\documentclass[12pt]{article}
\usepackage{amssymb}
\usepackage{epsfig}

\catcode`\@=11
\def\numberbysection{\@addtoreset{equation}{section}
        \def\theequation{\thesection.\arabic{equation}}}

\def\beq{\begin{equation}}
\def\eeq{\end{equation}}
\numberbysection
\begin{document}
\begin{titlepage}
\begin{center}
\hfill  \\
\vskip 1.in {\Large \bf Algebraic method for the harmonic oscillator with a minimal length} \vskip 0.5in P. Valtancoli
\\[.2in]
{\em Dipartimento di Fisica, Polo Scientifico Universit\'a di Firenze \\
and INFN, Sezione di Firenze (Italy)\\
Via G. Sansone 1, 50019 Sesto Fiorentino, Italy}
\end{center}
\vskip .5in
\begin{abstract}
We show that the algebraic method solving the ordinary harmonic oscillator can be adapted to the non-commutative case.
\end{abstract}
\medskip
\end{titlepage}
\pagenumbering{arabic}
\section{Introduction}

It has been suggested that a new interesting structure at short distances, characterized by a finite minimal position uncertainty $\Delta x_0$, can be modeled with a small modification of the commutation rules. It is also expected by plausibility arguments of string theory that such a short distance behaviour emerges at the Planck scale.

Moreover there are many examples from other physical problems involving quanta that cannot be localized to a point, i.e. quasi-particles and collective excitations. The most elementary example of a non point-like particle is the harmonic oscillator with a minimal length. This problem has been successfully solved in \cite{1}, using the momentum representation. In \cite{2} it has been discussed that for this case it is not possible to define generalized raising and lowering operators.

It is aim of this article to show instead that it is possible to solve this non-commutative problem with a totally algebraic method, based only on the modified commutation rule

\beq [ \hat{x}, \hat{p} ] \ = \ i \hbar \ ( 1 \ + \ \beta \ \hat{p}^2 \ ) \label{11} \eeq

This allows us to represent the excited states by applying a string of operators to the ground state, as one usually does in the standard case. This representation is useful to
solve the eigenvalue problem, compute all the normalizations and the expectation values of the operator $\hat{x}^2$ on the excited states.

\section{Ladder operators}

In literature it has been solved the quantization of the harmonic oscillator

\beq \hat{H} \ = \ \frac{1}{2m} \ ( \ \hat{p}^2 \ + \ m^2 \omega^2 \hat{x}^2 \ ) \ \label{21}\eeq

in presence of the commutation rule

\beq \ [ \ \hat{x}, \hat{p} \ ] \ = \ i \hbar \ ( \ 1 \ + \ \beta \hat{p}^2 \ ) \ \label{22} \eeq

The method employed is finding the eigenvalues and eigenvectors of the Hamiltonian (\ref{21}) in the momentum representation. Instead in the present article we want to solve the same problem using purely algebraic manipulations, based only on the commutation rule (\ref{22}).

In the case of the standard harmonic oscillator the quantization leads to equally spaced energy levels. Therefore it is possible to shift from one energy level to another by adding or removing a certain amount of energy quanta. This concept is practically realized by moving from the operators $\hat{x}$ and $\hat{p}$ to two news operators factorizing the Hamiltonian.
If $\hat{x}$ and $\hat{p}$ were commuting , then the Hamiltonian could be written as the product of two factors:

\beq \frac{1}{2m} \ ( \ - i \hat{p} \ + \ m \omega \hat{x} \ ) ( \ i \hat{p} \ + \ m \omega \hat{x} \ ) \label{23} \eeq

In presence of the standard commutation rule of quantum mechanics we can still write the Hamiltonian in a factorized form apart from a constant:

\beq \hat{H} \ = \ \hbar \omega \ ( \ a^{\dagger} a \ + \ \frac{1}{2} \ ) \label{24} \eeq

where we define the ladder operator

\beq a \ = \ \frac{ m \omega \hat{x} \ + \ i \hat{p} }{ \sqrt{ 2 m \hbar \omega }} \label{25} \eeq

In presence of the more complex commutation rule (\ref{22}), it is intriguing that it is again possible to write the Hamiltonian in a factorized form apart from a constant

\beq \hat{H} \ = \ \frac{\hbar \omega}{2} \ \left[ \ \frac{a^\dagger (\mu) a(\mu) \ + \mu }{
\sqrt{ \mu ( \mu-1 )} }
 \right] \label{26} \eeq

where

 \beq a(\mu) \ = \ \frac{ \hat{x} \ + \i \hbar \beta \mu \ \hat{p} }{\hbar \sqrt{\beta}}
 \ \ \ \ \ \ \ \ \ \ \mu \ = \ \frac{1}{2} \ + \ \frac{1}{2} \ \sqrt{ 1 + \frac{4}{m^2 \omega^2 \hbar^2 \beta^2} }
 \label{27} \eeq

and the parameter $\mu$ is fixed from the commutation rule (\ref{22}) and from the form of the Hamiltonian (\ref{21}). This factorization will be useful in the following to describe the construction of the ground state.

In the limit $\beta\rightarrow 0$ i.e. $\mu \rightarrow \infty$

 \beq a(\mu) \ \rightarrow \ \sqrt{2\mu} \ a \label{28} \eeq

where $a$ is the typical ladder operator of the harmonic oscillator, and the Hamiltonian
reduces to the standard form

 \beq H \ \rightarrow \ \hbar \omega \ \left( \ a^\dagger a \ + \ \frac{1}{2} \right)
 \label{29} \eeq

 \section{Commutation rules}

To show that the ladder operators $a(\mu)$, $a^\dagger(\mu)$ are useful to raise or lower the number of energy quanta it is necessary to study the commutation rule between $a(\mu)$ and $a^\dagger (\mu)$. In the case of the standard harmonic oscillator it is well known that

 \beq \ \ [ a, a^\dagger ] \ = \ 1 \label{31} \eeq

Unfortunately in our case the operators $a(\mu)$, $a^\dagger(\mu)$ do not form any closed algebra, however by carefully manipulating the possible combinations between $\hat{x}$ and $\hat{p}$ we have been able to derive the following meaningful rule

 \beq a(\mu) \ a^\dagger (\mu ) \ - \ a^\dagger (\mu+1) \ a(\mu+1) \ = \ 1 \ + \ 2\mu \label{32} \eeq

 We now apply the Hamiltonian operator in the form (\ref{26}) to $a^\dagger
 (\mu)$

 \beq H \ a^\dagger (\mu) \ = \ \frac{ \hbar\omega }{2} \left[ \frac{ a^\dagger (\mu) \ a^\dagger (\mu+1) \ a(\mu+1) \ + \ ( \ 1 \ + \ 3\mu \ ) \ a^\dagger (\mu) }{ \sqrt{\mu (\mu-1)} }  \right] \label{33} \eeq

This formula will be useful to discuss the construction of all higher energy states.

 \section{Spectrum of the Hamiltonian}

By applying the factorized form of the Hamiltonian (\ref{26}) to a generic state $\psi (p)$ we note that the mean value of energy reaches its minimum with the choice

 \beq a(\mu) \ \psi_0(\mu) \ = \ 0 \label{41} \eeq

Since the operator $\hat{x}$ is described in the momentum representation by

 \beq \hat{x} \ = \ i \ \hbar \ ( \ 1 + \ \beta p^2 \ ) \ \frac{\partial}{\partial p}
 \label{42} \eeq

the eigenfunction of the ground state is given by

 \beq \psi_0 \ ( \ \mu \ ) \ = \ \left( \sqrt{\frac{\beta}{\pi}} \ \frac{ \Gamma(\mu+1)}{\Gamma(\mu+\frac{1}{2})} \right)^{\frac{1}{2}} \ {( \ 1 \ + \ \beta \ p^2 \ )}^{-\frac{\mu}{2}}
\label{43} \eeq

and the corresponding eigenvalue is

\beq E_0 \ = \ \frac{\hbar\omega}{2} \ \sqrt{\frac{\mu}{\mu-1} } \label{44} \eeq

To build the first excited state it is enough to look at the equation (\ref{33}). To cancel the first contribution it is convenient to apply this relation to $\psi_0 (\mu+1)$

\begin{eqnarray}
a(\mu+1) \ \psi_0 (\mu+1) \ & = & \ 0 \nonumber \\
H  \ a^\dagger (\mu) \ \psi_0 (\mu+1) \ & = & \ \frac{\hbar\omega}{2} \ \frac{ \ 1 \ + \ 3\mu}{\sqrt{\mu (\mu-1) } } \ a^\dagger (\mu) \ \psi_0 ( \mu+1 ) \label{45} \end{eqnarray}

This formula allows us to define the first excited state as

\beq \psi_1 (\mu) \ = \ c_1 (\mu) \ a^\dagger (\mu) \ \psi_0 (\mu+1) \label{46} \eeq

with eigenvalue

\beq E_1 \ = \ \frac{\hbar\omega}{2} \ \frac{ \ 1 \ + \ 3\mu}{\sqrt{\mu (\mu-1) } } \label{47} \eeq

Given the structure of the commutation rule (\ref{32}) it is clear how to generalize to the case of the level $n$ excited state

\begin{eqnarray}
\psi_n (\mu) & = & c_n (\mu) \ a^\dagger (\mu) \ a^\dagger (\mu+1) \ ... \ a^\dagger (\mu \ + \ n \ - \ 1 \ ) \ \psi_0 \ ( \ \mu \ + n \ ) \ = \nonumber \\
& = & \frac{c_n (\mu) }{c_{n-1} (\mu+1) } \ a^\dagger ( \mu ) \ \psi_{n-1} (\mu+1) \label{48}\end{eqnarray}

This state satisfies to the condition

\beq a (\mu+n) \ ... \ a (\mu+1) \ a (\mu) \ \psi_n (\mu) \ = \ 0 \label{49} \eeq

It is easy to show that these states are all eigenvectors of the Hamiltonian (\ref{26}). By induction supposing that

\beq a^\dagger (\mu) \ a (\mu) \ \psi_n (\mu) \ = \ f(n,\mu) \ \psi_n (\mu), \label{410} \eeq

using the definition of the generic excited state (\ref{48}) and applying the rule (\ref{32}) we obtain the recurrence equation

\beq f(n,\mu ) \ = \ f(n-1,\mu+1) \ + \ 1 \ + \ 2\mu \label{411} \eeq

which is solved by

\beq f(n,\mu) \ = \ \sum_{i=0}^{n-1} \ ( \ 1 \ + \ 2 ( \ \mu \ + \ i ) \ ) \ = \ n^2 \ + \ 2 n \mu \label{412} \eeq

Therefore we conclude that the energy eigenvalues of the generic $n$-eigenstate are

\beq E_n \ ( \mu ) \ = \ \frac{\hbar\omega}{2} \ \left( \ \frac{ n^2 \ + \ 2 n \mu \ + \mu }{ \sqrt{\mu (\mu-1)} } \right) \ \ \ \ \ \rightarrow \ \ \ \hbar \omega \ ( \ n \ + \ \frac{1}{2} \ ) \ \ \ \ \ \  {\rm for } \ \mu \rightarrow \infty \label{413} \eeq

We have still to compute the undetermined coefficients $c_n (\mu)$. By imposing the orthogonality condition and applying several times the rule (\ref{32}) we obtain:

 \beq c_n (\mu ) \ = \ \left\{  \prod_{k=0}^{n-1} \ [ ( n-k )^2 + 2 ( \mu + k )( n-k ) ]
 \right\}^{-\frac{1}{2}} \label{414} \eeq

and finally

 \beq \frac{ c_n (\mu) }{ c_{n-1} (\mu+1) } \ = \ \frac{1}{\sqrt{ n^2 + 2 n \mu }}
 \label{415} \eeq

 \section{Comparison with the standard method}

The direct solution of the eigenvalue problem \cite{1}-\cite{3} gives rise to the following polynomial

 \beq {}_2 F_1 \ \left( \ - n \ , \ 2 s \ + \ 1 \ - \ n \ , \ s \ + \ 1 \ - \ n \ , \ \frac{1+ i \sqrt{\beta} p }{2} \ \right) \label{51} \eeq

  with the parameter $ s= - \mu $. It is well known in literature that these are nothing else than Jacobi polynomials

 \beq {}_2 F_1 \ ( \ - \ n \ , \ \alpha \ + \ 1 \ + \ \beta \ + \ n \ , \ \alpha \ + \
 1 ; x ) \ = \ \frac{ n! }{ ( \alpha + 1 )_n } \ P_n^{\alpha, \beta} \ ( 1-2x ) \label{52} \eeq

defined through the following formula

 \beq P_n^{\alpha,\beta} (z) \ =  \ \frac{ \Gamma \ ( \ \alpha \ + \ n \ + \ 1 )}{ n! \ \Gamma( \ \alpha \ + \ \beta \ + \ n \ + \ 1 \ ) } \ \sum^{n}_{m = 0}
 \left( \begin{array}{c} n \\ m \end{array} \right) \ \frac{ \ \Gamma \ ( \ \alpha \ + \ \beta \ + \ n \ + \ m \ + \ 1 \ ) }{ \ \Gamma \ ( \ \alpha \ + \ m \ + \ 1 \ )} \
 \left( \frac{z-1}{2} \right)^m \label{53} \eeq

They satisfy to the orthogonality condition:

\beq \int^1_{-1} \ P_m^{(\alpha,\beta)} \ (x) \ P_n^{(\alpha,\beta)} \ (x) \ ( \ 1 \ - \ x \ )^\alpha \ ( \ 1 \ + x \ )^\beta \ dx \ = \ c_n \delta_{n, m} \label{54} \eeq

where the coefficients $c_n$ are given by

\beq c_n \ = \ \frac{ 2^{ \ \alpha \ + \ \beta \ + \ 1} }{ 2n \ + \ \alpha \ + \ \beta \ + \ 1 \ } \ \frac{ \Gamma \ ( \ n \ + \ \alpha \ + \ 1 \ ) \ \Gamma \ ( \ n \ + \ \beta \ + \ 1 \ ) }{ n! \ \Gamma \ ( \ n \ + \ \alpha \ + \ \beta \ + \ 1 \ ) } \label{55} \eeq

The first polynomials are

\begin{eqnarray}
P_0^{\alpha,\beta} \ ( \ x \ ) & = & 1 \nonumber \\
P_1^{\alpha,\beta} \ ( \ x \ ) & = & \ \frac{1}{2} \ [ \ 2 \ ( \ \alpha \ + \ 1 \ ) \ + \ ( \ \alpha \ + \ \beta \ + \ 2 \ ) \ ( \ x \ - \ 1 \ ) \  ] \nonumber \\
P_2^{\alpha,\beta} \ ( \ x \ ) & = & \ \frac{1}{8} \ [ \ 4 \ ( \ \alpha \ + \ 1 )( \ \alpha \ + \ 2\ ) \ + \ 4 \ ( \ \alpha \ + \ \beta \ + \ 3 \ )( \ \alpha \ + \ 2 \ )( \ x \ - \ 1 \ ) \ + \ \nonumber \\
& + & \ ( \ \alpha \ + \ \beta \ + \ 3 \ )( \ \alpha \ + \ \beta \ + \ 4 \ ) ( \ x \ - \ 1 \ )^2 ] \label{56} \end{eqnarray}

In our problem we notice that the parameters $\alpha$ and $\beta$ are fixed to be equal to

\beq \alpha \ = \ \beta \ = \ - \ \mu \ - \ n \label{57} \eeq

For example the wave function of the second excited state is proportional to

\beq P_2^{(-\mu-2,-\mu-2)} \ ( - i \sqrt{\beta} p ) \ \propto \ ( \ 1 \ - \ ( \ 1 \ + \ 2 \ \mu \ ) \ \beta \ p^2 \ ) \ \label{58} \eeq

Comparing with our method we can give the representation of this state as

\beq \psi_2(\mu) \ \sim \ a^\dagger \ ( \mu ) \ \psi_1 \ ( \ \mu \ + \ 1 \ ) \ \sim \
a^\dagger \ ( \mu ) \ p \ ( \ 1 \ + \ \beta \ p^2 \ )^{ - \ \frac{ \mu \ + \ 2 \ }{2}} \label{59} \eeq

obtaining exactly the same polynomial defined above.

\section{Computation of the expectations values}

Since the excited states are obtained applying the string of operators $a^\dagger (\mu)$
that have definite parity with respect to the exchange $p \rightarrow -p$,
we suddenly observe that the following expectation values

\beq < \psi_n (\mu) | \ \hat{x} \ | \psi_n (\mu) > \ = \ < \psi_n (\mu) | \ \hat{p} \ | \psi_n (\mu) > \ = \ 0 \label{61} \eeq

are zero as obtained with the standard quantization.

The more involved case is the expectation value of the operator $\hat{x}^2$. To reach this aim we observe that

\begin{eqnarray}
| \psi_n (\mu) > & = & \frac{ a^\dagger ( \mu ) }{ \sqrt{ \ n^2 \ + \ 2 \ n \ \mu } } \ | \psi_{n-1} \ ( \ \mu \ + \ 1 ) > \nonumber \\
\hat{x} \ | \psi_n (\mu) > & = & \frac{ \hbar\sqrt{\beta}}{2} \ [ \ a^\dagger (\mu) \
| \psi_n (\mu) > \ + \ \sqrt{ \ n^2 \ + \ 2  n \mu } \ | \psi_{n-1} ( \mu \ + \ 1)) > \ ] \label{62} \end{eqnarray}

and the matrix elements of $\hat{x}^2$ is

\begin{eqnarray}
< \psi_n (\mu) | \ \hat{x}^2 \ | \psi_n (\mu) > & = & \frac{ \hbar^2 \beta }{4} \ [ \
\psi_n (\mu) | \ a(\mu) \ a^\dagger (\mu) \ | \psi_n (\mu) > \ + \ \nonumber \\
& + & < \psi_{n-1} ( \mu+1 ) | \ a^2 (\mu) \ + \ a^{\dagger 2} (\mu) \ | \psi_{n-1} (\mu+1) > \ + \ \nonumber \\
& + & ( \ n^2 \ + \ 2 \ n \ \mu \ ) \ ] \label{63} \end{eqnarray}

Since

\begin{eqnarray}
 \frac{\hbar^2 \beta}{4} \ a(\mu) \ a^\dagger (\mu) & = & \frac{1}{4} \ ( \ \hat{x}^2 \ + \ \hbar^2 \beta^2 \mu ( \mu+1 ) \ \hat{p}^2 \ + \ \hbar^2 \beta \mu \ ) \nonumber \\
\frac{\hbar^2 \beta}{4} \ ( \ a^2 (\mu) \ + \ a^{\dagger 2} (\mu) \ ) & = &
\frac{1}{2} ( \ \hat{x}^2 \ - \ \hbar^2 \beta^2 \mu^2 \ \hat{p}^2 \ )
\label{64}, \end{eqnarray}

defining

\beq < \psi_n (\mu) | \ \hat{x}^2 \ | \psi_n(\mu) > \ = \ f( n,\mu ) \ \ \ \ \ \
< \psi_n (\mu) | \ \hat{p}^2 \ | \psi_n(\mu) > \ = \ g( n,\mu ) \label{65} \eeq

we arrive at the conclusion that

\begin{eqnarray}
3 f(n,\mu) & = & \ \hbar^2 \beta^2 \mu ( \mu +1 ) \ g(n,\mu) \ + \ 2 \ [ \ f(n-1,\mu+1) \ - \ \hbar^2 \beta^2 \mu^2 \ g(n-1, \mu+1) \ ] \ + \nonumber \\\ &  +
&  \hbar^2 \beta \ ( \ n^2 \ + \ 2 n \mu \ + \ \mu \ ) \label{66} \end{eqnarray}

We can add to this equation the information due to the eigenvalue equation

\beq f(n,\mu) \ + \ \hbar^2 \beta^2 \mu (\mu-1) \ g(n,\mu) \ = \ \hbar^2 \beta \ ( \
n^2 \ + \ 2 n \mu \ + \ \mu \ ) \label{67} \eeq

By taking the difference between the last two equations we obtain

\beq f(n,\mu) \ = \ \hbar^2 \beta^2 \mu^2 \ g(n,\mu) \ + \ f(n-1,\mu+1) \ - \ \hbar^2\beta^2\mu^2 \ g(n-1,\mu+1) \label{68} \eeq

that, together with the eigenvalue equation (\ref{67}), gives rise to the recurrence equation

\beq g(n,\mu) \ = \ \frac{2\mu+1}{2\mu-1} \ g(n-1,\mu+1) \ + \ \frac{2}{\beta \ ( 2 \mu-1 )} \label{69} \eeq

This is solved simply by the following result

\begin{eqnarray}
g(n,\mu) & = & \frac{2n+1}{\beta (2\mu-1)} \nonumber \\
f(n,\mu) & = & \hbar^2 \beta \ \left( \ n^2 \ + \ ( 2n+1 ) \ \frac{\mu^2}{2\mu-1} \right)
\label{610} \end{eqnarray}

As you can see, many of the manipulations that one usually does with the ladder operators
can be extended to the rule (\ref{32}).

As a final note we can add the construction of the coherent states. These can be defined
as the eigenfunctions of the ladder operator $a(\mu)$:

\beq a(\mu) \ \psi_{\lambda} (\mu) \ = \ \frac{\lambda}{\hbar \ \sqrt{\beta}}  \ \psi_{\lambda} (\mu)  \label{611}
\eeq

This equation is solved by

\beq \psi_\lambda \ (\mu) \ \sim ( \ 1 \ + \ \beta \ p^2 \ )^{-\frac{\mu}{2}} \
e^{-\frac{i \lambda}{\hbar \sqrt{\beta}} \ \arctan
\sqrt{\beta p^2} } \label{612} \eeq

These states are analogous to the maximally localized states studied in \cite{1}.

\section{Conclusions}

We have learned from this article that it is possible in a non-commutative geometry to represent the eigenfunctions of the harmonic oscillator in a operatorial form. The major
obstacle we have found is that the commutation rules between the raising and lowering operators are non-standard, i.e. they do not form a closed algebra. However one can still perform many of the manipulations of quantum mechanics like reproducing the results obtained with the differential equation method and computing
the expectation values of the operators $\hat{x}^2$ e $\hat{p}^2$.
Moreover it is possible define a non-commutative version of the coherent states, that look like the maximally localized states studied in \cite{1}.

Finally we have tried to generalize this operatorial method to the quantization of many harmonic oscillators with a minimal length \cite{4}-\cite{5}, finding no interesting structure.
However we will discuss this problem in detail in a forthcoming paper.

\end{document}